\documentclass[11pt,twoside]{article}
\usepackage{asp2010}

\resetcounters

\begin{document}

\title{Helioseismic and Magnetic Imager Multi-height Dopplergrams}

\author{Kaori~Nagashima,$^1$ Laurent Gizon,$^{1,2}$ Aaron Birch,$^1$  
Bj\"orn L\"optien,$^{2}$ Sebastien Couvidat,\altaffilmark{3} and
Bernhard Fleck\altaffilmark{4}
\affil{$^1$Max-Planck-Institut f\"ur Sonnensystemforschung, 37191 Katlenburg-Lindau, Germany}
\affil{$^2$Institut f\"ur Astrophysik, Georg-August-Universit\"at, 37077 G\"ottingen, Germany}
\affil{$^3$HEPL, Stanford University, Stanford, CA 94305, USA}
\affil{$^4$ESA Science Operations Department, c/o NASA/GSFC, Greenbelt, MD 20771, USA}}

\begin{abstract}
We study Doppler velocity measurements at multiple 
heights in the solar atmosphere using a set of six filtergrams obtained by the Helioseismic magnetic Imager on board the Solar Dynamics Observatory.
There are clear and significant phase differences between core and wing Dopplergrams
in the frequency range above the photospheric acoustic cutoff frequency, which
indicates that these are really ``multi-height" datasets.  
\end{abstract}

\section{Background}
In recent helioseismology studies, photospheric Dopplergrams (i.e., maps of the line-of-sight velocity of the photosphere) obtained by the Helioseismic and Magnetic Imager 
\citep[HMI;][]{2012SoPh..275..229S} 
on board the Solar Dynamics Observatory \citep[SDO;][]{2012SoPh..275....3P} 
have been used to investigate the solar interior.
Multi-height velocity information from observations 
would be very useful not only for helioseismology analyses \citep[e.g.,][]{2009ApJ...694L.115N}, but also for other purposes,
in particular for studies of the energy transport by waves in the solar atmosphere 
\citep[e.g.,][]{2006ApJ...648L.151J, 2008ApJ...681L.125S,2009ASPC..415...95S,
2011A&A...532A.111K,2010ApJ...723L.134B}.

\section{HMI observation datasets}
HMI takes filtergrams at 6 wavelengths 
(+172.0 m\AA\ ($I_0$), +103.2 m\AA\  ($I_1$),
 +34.4 m\AA\ ($I_2$), -34.4 m\AA\  ($I_3$), -103.2 m\AA\  ($I_4$), 
and -172.0 m\AA\  ($I_5$) ) around the Fe \textsc{i} absorption line at 6173 \AA . Standard Dopplergrams provided by the HMI pipeline \citep{2012SoPh..278..217C} are derived from these six filtergrams.
The formation height of the HMI Doppler signal is estimated
to be about 100 km above the $\tau_{\mathrm{5000 \AA}} = 1$ surface \citep{2011SoPh..271...27F}.
In this study, instead, we try to create 
multi-height Dopplergrams from the HMI filtergrams. 

The observed Fe \textsc{i} line profile shifts in wavelength 
because of the SDO orbital motion. 
Since SDO is in a geosynchronous orbit, the SDO orbital velocity toward the Sun is not constant and the absolute value of its line-of-sight (LOS) velocity towards the Sun reaches up to $\sim 3.5 \ \mathrm{km \ s^{-1}}$ \citep{2012SoPh..275..229S}.
Figure \ref{fig:obs_lineprof} shows some sample observed line profiles. 
To create these line profiles we averaged
the intensity value at each wavelength  
over a field of view  of $30 \ \mathrm{degrees}$ square, and hence,
the line shift is basically caused by the observer motion. 
The vertical dotted line indicates the line shift calculated from the observer LOS velocity in the data headers. 
We use this observer motion to calibrate the multi-height Dopplergrams.

\articlefigure[scale=0.9]{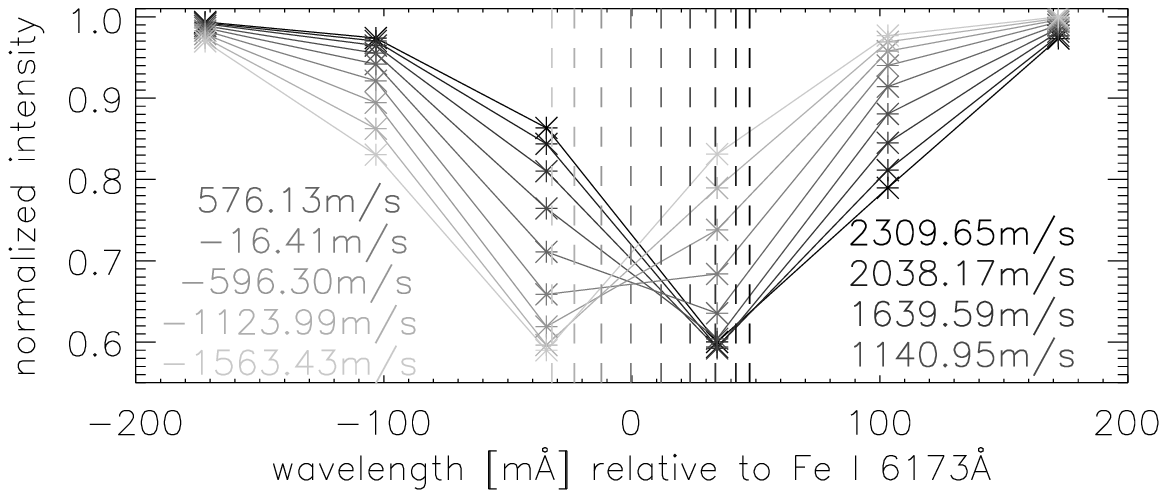}{fig:obs_lineprof}
{Sample SDO/HMI observed line profiles. Different gray scales indicate different times, and the numbers on the panel are the observer LOS velocities at these times.
The line shift is caused by the spacecraft velocity. 
The vertical dotted line indicates the line shift calculated from 
the observer LOS velocity in the data headers.}

\section{Multi-height Dopplergrams}
We calculate three Doppler signals using the set of six filtergrams: 
$D_{\mathrm{core}} \equiv (I_3 -I_2)/(I_3+I_2)$,
$D_{\mathrm{wing}} \equiv (I_4 -I_1)/(I_4+I_1)$, and 
$D_{\mathrm{far-wing}} \equiv (I_5-I_0)/(I_5+I_0)$,
and convert them into the Doppler velocity, 
$V_{\mathrm{core}} =f_{\mathrm{core}}(D_{\mathrm{core}})$,
$V_{\mathrm{wing}} =f_{\mathrm{wing}}(D_{\mathrm{wing}})$, 
$V_{\mathrm{far-wing}} =f_{\mathrm{far-wing}}(D_{\mathrm{far-wing}})$.
To get the conversion formula, we use the spatial average Doppler signal over 
a 30-degree-square area near the disc center and the LOS component of 
SDO observer motion. The functions $f(D)$ are 
defined by 3rd-order polynomial fitting and the 
fitting parameters are calculated from three-day observation datasets 
 as shown in Figure \ref{fig:obs_fitting}.
Here we use six non-overlapping nine-hour datasets obtained between 
January 22 0UT and January 24 15UT, 2011, except that
we exclude the period from January 22 18UT to January 23 3UT because 
it has a long data gap.
We track 30-degree-square quiet-Sun regions at the Carrington rate for nine hours each
using mtrack \citep{2011JPhCS.271a2008B}.
For each run, the central point of the field of view passes the disc center at the mid-point 
of the run. 

This method to convert the Doppler signal $D$ into 
the Doppler velocity $V$ is limited within a certain velocity range for two reasons:
1) If the velocity is too large, the Doppler shift of the line is too big and 
the line center is outside of the blue and red pairs.
For the core ($I_2$ and $I_3$) this limitation is severe; 
if the velocity exceeds 1.7 km \ s$^{-1}$
the line center is outside the blue and red pair ($\pm 34.4$ m\AA). 
2) Since the SDO motion is less than $\sim 3.5$ km \ s $^{-1}$,
the fitting is limited within the range. 

\articlefigure[]{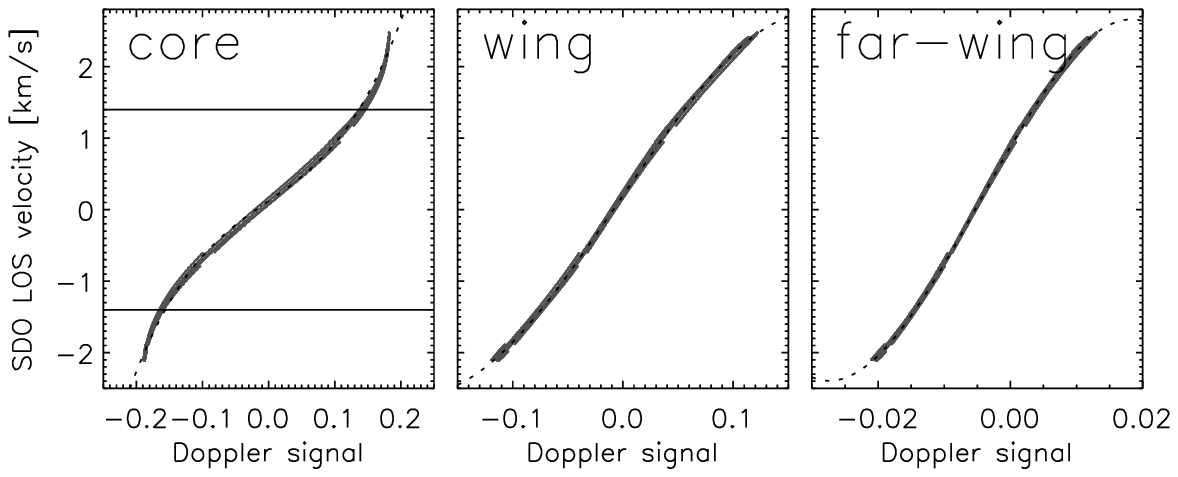}{fig:obs_fitting}{
Scatter plots of mean Doppler signal and the observer LOS velocity.
The dotted curves are third-order polynomials and obtained by least-square 
fitting of the data points.
For the core, if the velocity exceeds $1.4\  \mathrm{km  \ s^{-1}}$ (solid horizontal lines)
the data points are not used for the fitting.}

\section{Do these Doppler shifts really correspond to velocities at multiple heights?}

\begin{figure}[tbh]
\plotone{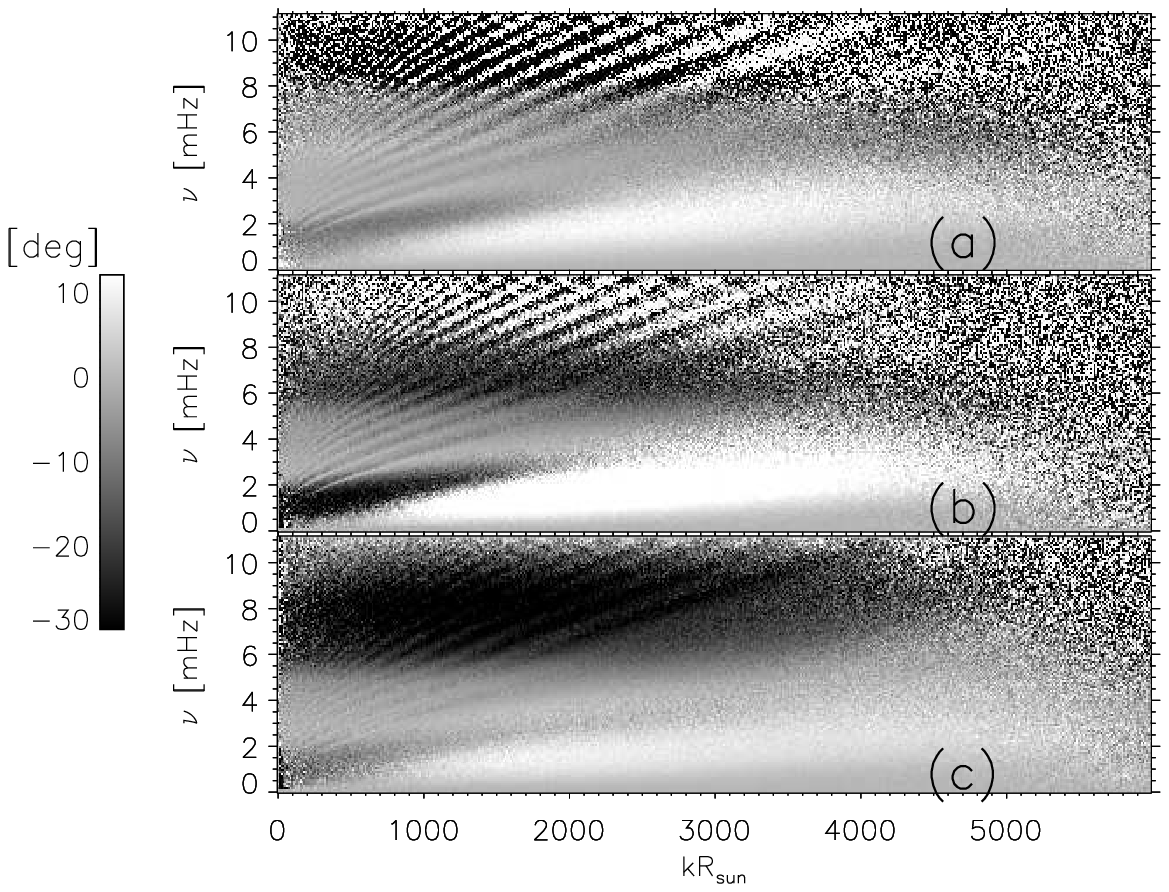}
\caption{Phase differences: (a)$\phi_{\mathrm{Far-wing}}-\phi_{\mathrm{Wing}}$,
(b) $\phi_{\mathrm{Far-wing}}-\phi_{\mathrm{Core}}$,
(c) $\phi_{\mathrm{Wing}}-\phi_{\mathrm{Core}}$.
Clear phase differences in the frequency range above the 
acoustic cutoff frequency are seen in (c). \label{fig:phasedif}}
\end{figure}

Correlation coefficients between the Doppler velocities of one snapshot
at the time of minimum SDO motion speed on January 23, 2011,
are 0.95 ($V_{\mathrm{core}}$ and $V_{\mathrm{wing}}$),
0.83 ($V_{\mathrm{far-wing}}$ and $V_{\mathrm{wing}}$),
and 0.76 ($V_{\mathrm{far-wing}}$ and $V_{\mathrm{core}}$). 
The phase differences between pairs of the Dopplergrams, however, 
show that $V_{\mathrm{core}}$, $V_{\mathrm{wing}}$, and $V_{\mathrm{far-wing}}$
have promise as multi-height Dopplergrams.
Figure \ref{fig:phasedif} shows the phase differences among the three Doppler velocities.
Figure \ref{fig:phasedif}(c) shows clear phase differences
between the wing and core
in the frequency range above the photospheric acoustic cutoff frequency 
($\sim 5.4 $ mHz). 
The phases referred to the far-wing are noisier (panels (a) and (b)); this might be because the far-wing intensity level is very close to the continuum level, and small fluctuations of the intensity ($I_0$ or $I_5$) might cause large velocity differences. 
In the g-mode area (with low frequency and large wavenumber), 
the sign of the phase difference is opposite to what we have
in the higher frequency range; this might be a signature of atmospheric 
gravity waves, which is consistent with e.g, 
\citet{2008ApJ...681L.125S, 2009ASPC..415...95S}.
This phase difference also indicates that the 
two velocities are formed in different heights in the atmosphere.

\citet{2008A&A...481L...1M} used photospheric and 
chromospheric intensity datasets, instead of Dopplergrams,
obtained by the Solar Optical Telescope on board 
the Hinode satellite.
The weak phase shifts above the acoustic cutoff 
frequency and strong phase shifts in the g-mode area
shown in their Figure 1 are consistent with ours,
while the strong phase difference along the p-mode ridges which they reported 
is not seen in our results. 
The interpretation of the intensity datasets, however, 
might not be straightforward because of radiation effects. 
Phase differences between two velocity datasets
provide much clearer diagnostics, as they simply 
measure actual travel times between the two layers.

To estimate the heights of the contribution layers of the multi-height Dopplergrams 
we plan to use line synthesis calculations in a realistic solar atmospheric model
\citep{nagashima_prep}.

\acknowledgements
The HMI data used are courtesy of NASA/SDO and the HMI science team.
The German Data Center for SDO, funded by the German Aerospace Center (DLR), provided the IT infrastructure.
This work was carried out using the data from the SDO HMI/AIA Joint Science Operations Center Data Record Management System  and Storage Unit Management System (JSOC DRMS/SUMS).

\end{document}